\documentclass[aps,prd,showpacs,preprintnumbers,nofootnotebib]{revtex4}
\usepackage{graphics}
\usepackage{multirow}
\def \beq{\begin{equation}}
\def \eeq{\end{equation}}
\def \beqa{\begin{eqnarray}}
\def \eeqa{\end{eqnarray}}
\def \tr{{\rm Tr}\,}

\def \alphae{\alpha_{\scriptscriptstyle E}}
\def \alphas{\alpha_{\scriptscriptstyle S}}
\def \alphav{\alpha_{\scriptscriptstyle V}}
\def \alphams{\alpha_{\overline{\scriptscriptstyle MS}}}
\def \lms{\Lambda_{\overline{\scriptscriptstyle MS}}}
\def \L{\langle|L|\rangle}
\def \P{\langle P\rangle}
\def \chil{\chi_{\scriptscriptstyle L}}
\def \ie{{\sl i.e.\/}}
\def \etal{{\sl et al.\/}}
\def \jhep{{\sl J.\ H.\ E.\ P.\/}}

\def \np{{\sl Nucl.\ Phys.\/}}
\def \pl{{\sl Phys.\ Lett.\/}}
\def \pr{{\sl Phys.\ Rev.\/}}
\def \prl{{\sl Phys.\ Rev.\ Lett.\/}}

\begin{document}
\title{Scaling and the continuum limit of gluo$N_c$ plasmas}
\author{Saumen \surname{Datta}}
\email{saumen@theory.tifr.res.in}
\affiliation{Department of Theoretical Physics, Tata Institute of Fundamental
         Research,\\ Homi Bhabha Road, Mumbai 400005, India.}
\author{Sourendu \surname{Gupta}}
\email{sgupta@tifr.res.in}
\affiliation{Department of Theoretical Physics, Tata Institute of Fundamental
         Research,\\ Homi Bhabha Road, Mumbai 400005, India.}

\begin{abstract}
We investigated the finite temperature ($T$) phase transition for
SU($N_c$) gauge theory with $N_c=4$, 6, 8 and 10 at lattice spacing,
$a$, of $1/(6T)$ or less.  We checked that these theories have first order
transitions at such small $a$. In many cases we were able to find
the critical couplings with precision as good as a few parts in $10^4$. We
also investigated the use of two-loop renormalization group equations
in extrapolating the lattice results to the continuum, thus fixing
the temperature scale in units of the phase transition temperature,
$T_c$. We found that when $a\le1/(8T_c)$ the two-loop extrapolation
was accurate to about 1--2\%.  However, we found that trading $T_c$
for the QCD scale, $\lms$, increases uncertainties significantly, to the
level of about 5--10\%.
\end{abstract}
\preprint{TIFR/TH/09-34,hep-lat/yymmnnn}
\maketitle

\section{Introduction}
\label{sc.intro}

Since the realization \cite{thooft} that a non-trivial and tractable limit
is obtained for SU($N_c$) gauge theories when the gauge coupling, $g$,
is taken to zero and the number of colours, $N_c$, is taken to infinity,
keeping the combination $g^2 N_c$ fixed, there has been much work on this
limit \cite{catchall}. Most such work sums large classes of Feynman
diagrams and therefore is closely related to
perturbation theory. The hope is that
the limiting theory and a small number of corrections in a series in
$1/N_c$ would allow us to understand the physically interesting theory
with $N_c=3$. Lattice calculations are of help in testing this conjecture
by making the connection from the other direction--- by simulations
and complete non-perturbative computations at finite $N_c$. They test
whether a short series in $1/N_c$ for $N_c\ge3$ extrapolates correctly
to the tractable limit of $N_c\to\infty$. However, in order to test the
continuum computations, one must also take the continuum limit of the
lattice theories. This is the main thrust of this paper.

The theory with $N_c=3$ has been studied extensively before \cite{nc3},
and its continuum extrapolation using the renormalized weak coupling
expansion has been studied and found to work \cite{scaling}.  We shall
have occasion to use these results at various points in this paper.  The
finite temperature transition has been studied before in 3+1 dimensions
on lattices with $a=1/(4T)$ for $N_c=4$ \cite{nc4}. These early studies
found that the crossover from strong to weak coupling, which is a lattice
artifact, interfered with the finite temperature transition. Variant
actions were invented to solve this problem \cite{datta}. A modern
solution which depends on today's vastly improved computational power
is to just go to larger $N_t$ with the simplest action. For larger $N_c$
there have been some studies recently with $N_t=5$, 6 and 8 \cite{nclarger}.
These earlier works have presented evidence for a first-order thermal
phase transition.
For $N_c=4$, $\lms$ has been extracted from data on the string tension
in the Schr\"odinger functional scheme \cite{ncsf}.

In this paper we investigate the continuum limit of the finite temperature
deconfinement transition in SU($N_c$) pure gauge theory for $N_c>3$.
The main thrust of our study is to control the approach to the continuum
limit by performing simulations of the 3+1 dimensional theories at a
succession of lattice spacings, $a$, and then using the weak coupling
expansion for the extrapolation to zero lattice spacing. It turns out that
with today's computational power it is quite possible to reach lattice
spacings small enough for two-loop renormalization group equations (RGEs)
to be useful for the continuum extrapolation.  Indeed, at the lattice
spacings that we use, even the one-loop flow is a good rough indicator
of the continuum limit.

In order to perform these precision tests of the continuum limit
we performed lattice simulations of SU(4), SU(6), SU(8) and SU(10)
theories. In all cases we simulated theories with lattice cutoffs
of $a=1/(6T)$ and $1/(8T)$, and in some cases for even smaller
lattice spacings, going down to lattice spacing of $1/(12T)$ in one
case. We performed finite size scaling studies, thus extrapolating to
the thermodynamic limit of infinite spatial volumes, to check that
the thermal phase transitions is actually of first order at lattice
spacings $a\le1/(6T)$. Coupled with the continuum extrapolations that
we discuss next, this verifies earlier arguments about the order of the
finite temperature phase transition in continuum theories with $N_c\ge3$
\cite{svetitsky}.

Through the finite size scaling analysis we located the phase transition
point with a statistical precision of a few parts in $10^4$. We found
that the location of the phase transition point scales as expected in
the limit of $N_c\to\infty$.  With this precision we could test the
two-loop RG flow to a statistical accuracy of a few parts in $10^3$. It
turned out that at lattice spacing of $a\le1/(8T_c)$, the two-loop RGE
is trustworthy in extrapolation towards the continuum, within $3\sigma$
of the statistical accuracy. In all this the quantity $T_c$ is used to
set the scale of measurements.

Any test of a weak coupling expansion involves the choice of an RG scheme,
\ie, a choice of a measurement used to define the running coupling in the
gauge theory. If the perturbation theory is accurate, and all orders in
the expansion are available, then the choice of the scheme is immaterial
for any measurement. However, in all practical cases only a small number
of terms in the weak coupling expansion are available. We found that for
the determination of the temperature scale in terms of $T_c$ the scheme
dependence is statistically significant, but small in magnitude, being
around 1--2\%. This indicates that the lattice spacings used in
our study are small enough for the use of the weak-coupling expansion.
It seems likely that three-loop computations can improve matters.

This could be the first indication that non-perturbative lattice
computations for $N_c>3$ are at a point where they are more reliable than
the perturbative series needed for the continuum extrapolation.  Needless
to say, one could just push the non-perturbative lattice simulations to
smaller and smaller $a$ until the running coupling (at the scale of $a$)
decreases significantly and the available perturbative expansions begins
to be more accurate.  However, it is more cost-effective to develop
the perturbation theory to higher order.

In performing a weak coupling expansion the scale of choice is one which
defines how fast the coupling changes asymptotically when measured at
two different length scales. This intrinsic scale of QCD is called
$\lms$. We found that the determination of $\lms$ in terms of the
non-perturbatively determined scale $T_c$ is quite uncertain. While the
statistical errors are under control, the scheme dependence is quite
large. Our observations seem to indicate that one needs smaller lattice
spacings to stabilize the transformation from $T_c$ to $\lms$.

This paper is structured as follows: in the next section we discuss the
technicalities of the lattice simulations. Following this we present
our results for the finite temperature transition and its extrapolation
to the thermodynamic limit. Next, we discuss the continuum limit, the
setting of the temperature scale and the extraction of $\lms$. The final
section contains a summary of our results.
Some parts of our results have been
reported earlier in conference proceedings \cite{confs}.

\section{Simulations, measurements and other technicalities}
\label{sc.simul}

\begin{table}[htb!]
\begin{center}
\begin{tabular}{|r|cccc|}
\hline
$N_t$ & $N_c=4$ & $N_c=6$ & $N_c=8$ & $N_c=10$ \\
\hline
4 & 12, 16, 18, 20, 24 & 12, 16, 20 & & \\
6 & 16, 18, 20, 22, 24 & 14, 16, 18, 20, 24 & $16^*$ & $16^*$ \\
8 & 22, 24, 28, 30 & 20, $24^*$ & $16^*$ & $16^*$ \\
10 & 24 & $24^*$ & & \\
12 & 24 & & & \\
\hline
\end{tabular}
\end{center}
\caption{For each $N_c$ and $N_t$ the values of $N_s$ used in the
  simulations are given. Runs which are exploratory are marked by
  asterisks. The remaining runs are meant to yield precision data;
  for these the details of the statistics are given in
  Table \ref{tb.rewt}. Zero temperature runs were performed for
  $N_s=N_t=16$ for all $N_c$ and $N_s=N_t=24$ for $N_c=4$ and 6.}
\label{tb.runs}\end{table}

\begin{figure}
\begin{center}
   \scalebox{1.0}{\includegraphics{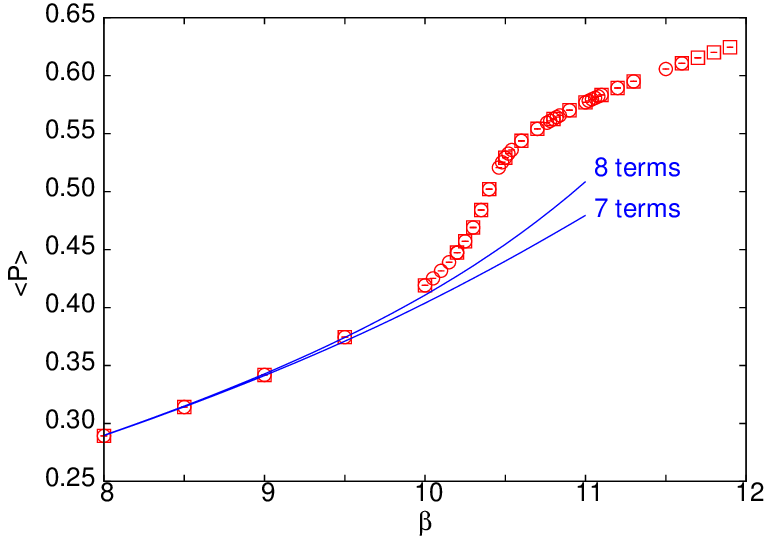}}
\end{center}
\caption{The average plaquette as a function of the bare coupling for
  $N_c=4$ on a $16^4$ lattice. Above $\beta=10$ the strong coupling
  series no longer predicts $\P$ accurately, and the
  theory crosses over to the weakly coupled phase.}
\label{fg.bulkco4}
\end{figure}

\begin{figure}
\begin{center}
   \scalebox{0.7}{\includegraphics{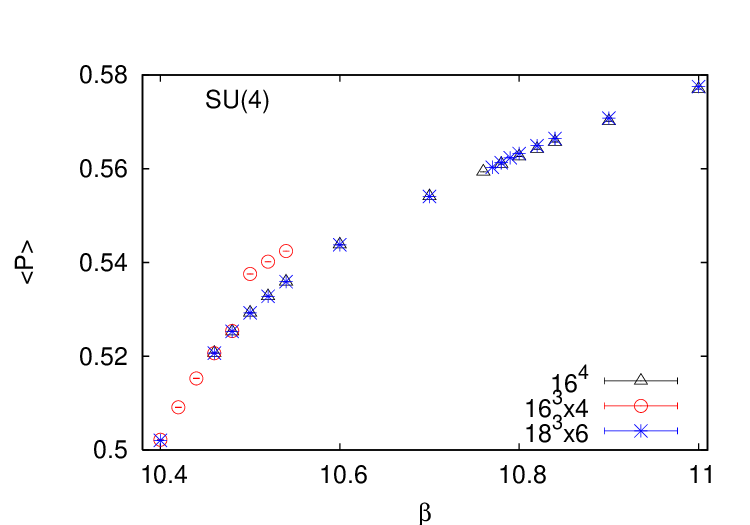}}
   \scalebox{0.7}{\includegraphics{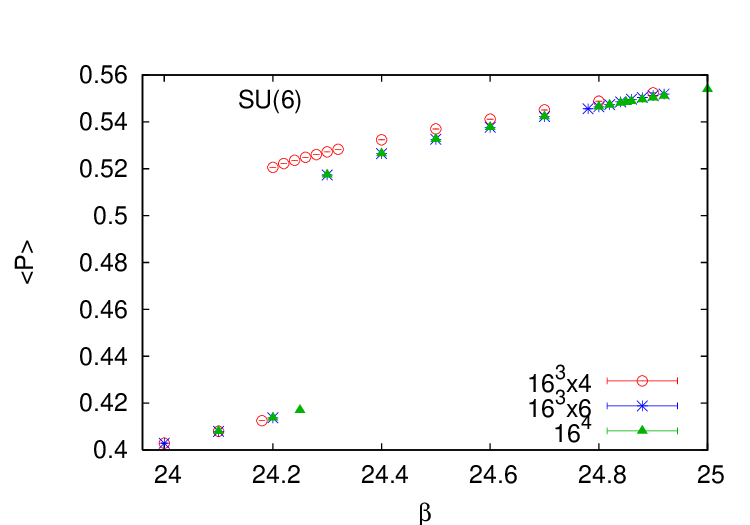}}
\end{center}
\caption{The average plaquette as a function of the bare coupling
  for $N_c=4$ and 6 on various lattice sizes. For SU(4) on a lattice with
  $N_t=4$ there is a jump in $\P$ at $\beta_c=10.48$ where
  the first order thermal phase transition occurs with a jump in $\L$.
  However, at larger $N_t$ there is no jump in $\P$. For
  SU(6) there is a jump in $\P$ at all $N_t$. For $N_t=4$
  the jump occurs at the thermal phase transition, but at all other $N_t$
  the bulk and thermal transitions are decoupled.}
\label{fg.bulk}
\end{figure}

\begin{figure}[b]
\begin{center}
   \scalebox{0.7}{\includegraphics{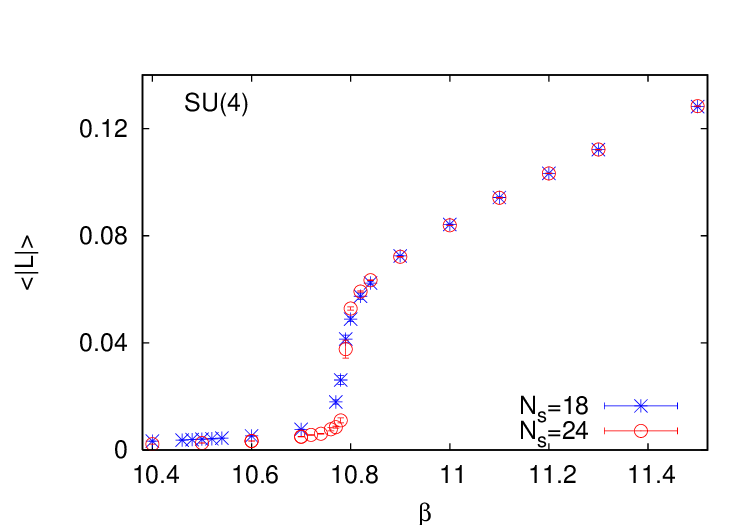}}
   \scalebox{0.7}{\includegraphics{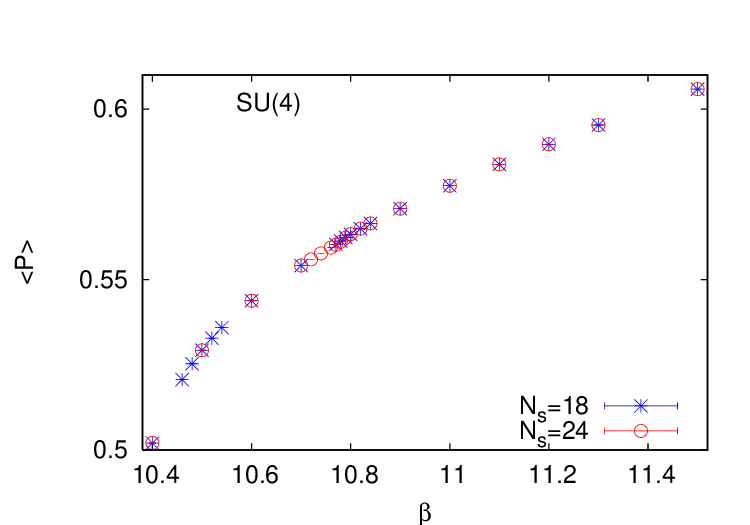}}
\end{center}
\caption{$\L$ and $\P$ at functions of $\beta$ on $6\times18^3$
  and $6\times24^3$ lattices for SU(4). A rapid change at $\beta_c=10.78$
  is seen in $\L$, whereas $\P$ is continuous. For
  $N_t>4$ the bulk and thermal transitions are decoupled for all $N_c$, as
  shown by this kind of observation.}
\label{fg.betac4}
\end{figure}

In this study we use the Wilson action---
\beq
   S = \beta\sum_{i,\mu<\nu} \left[1-{\rm Re}\,P_{\mu\nu}(i)\right],
\label{action}\eeq
where $P_{\mu\nu}(i)$ is the trace of the product of SU($N_c$) valued
link matrices, $U$, around a plaquette, starting from the
site $i$ and touching the site $i+\mu+\nu$. The trace is normalized
by a factor of $N_c$, so that by this definition the trace of an unit
matrix is unity. The lattices have size $N_t\times N_s^3$ in units of the
lattice spacing, $a$. The physical extent of the lattice is $aN_t=1/T$
and $\ell=aN_s=\zeta/T$ where $\zeta=N_s/N_t$ is called the aspect
ratio. Increasing $\zeta$ at fixed $T$ corresponds to increasing the
volume, $V=\ell^3$. The bare gauge coupling is $g^2=2N_c/\beta$.

The partition function,
\beq
   Z(V,T) = \int {\cal D}U {\rm e}^{-S[U]},
\label{partition}\eeq
is sampled using a Monte Carlo procedure in which over-relaxation steps
are mixed with heat-bath updates.  A large fraction of the CPU time is
taken up in the computation of the product of matrices connecting to
a given matrix (called staples). This computation scales as $N_c^3$,
since the time is dominated by the multiplication of $N_c\times N_c$
matrices. Therefore, for each computation of a staple, it would make
sense to update each of the $N_c(N_c-1)/2$ SU(2) subgroups of SU($N_c$)
a fixed number of times \cite{algo}. When we update all SU(2) subgroups
once in every step of a composite sweep which contains three steps of
over-relaxation per step of heat-bath, then about 50\% of the CPU time is
spent in the computation of staples, about 33\% in the over-relaxation
update, and about 12\% in the heat-bath. The rest of the time is spent
in the measurement of plaquettes and Polyakov loops. These fractions are
almost independent of $N_c$, whereas the actual CPU time per link update
scales very close to $N_c^3$.  It was argued earlier \cite{wolff} that in
an optimum hybrid over-relaxation algorithm the number of over-relaxation
steps should be increased linearly with $N_s$. If this were to be done,
then relatively less time would be spent in computing staples, resulting
in more optimal use of CPU time.

We performed simulations of four theories. An overview of the runs
is given in Table \ref{tb.runs} and its caption.  Almost all zero
temperature runs collected statistics of several tens of thousands of
composite sweeps, and most runs have statistics of over half a million
composite sweeps. The statistics of a set of measurements should actually
be judged by the auto-correlation time, $\tau_{int}$, since the error
in a measurement, $E$, is related to the variance of the measurements,
$\sigma^2$, through the formula $E^2 = \tau_{int}\sigma^2/N$ where $N$ is
the number of measurements. Auto-correlation functions of the plaquette
at $T=0$ show that $\tau_{int}$ varies between approximately 1 and 10
sweeps.  Since we study first order phase transitions, $\tau_{int}$ in
the transition region for the order parameter, $L$, is closely related
to the number of tunnelings between different phases \cite{fse}. The
statistics collected close to the transition region are summarized in
Table \ref{tb.rewt}.

SU($N_c$) theories with the action in eq.\ (\ref{action}) exhibit a bulk
transition when $N_c$ is large enough. This transition can be monitored
in zero temperature simulations using the expectation value of the
plaquette, $\P$, where
\beq
   P = \frac2{d(d-1)N_s^3N_t}\sum_{i,\mu<\nu}{\rm Re}\,P_{\mu,\nu}(i),
\label{plaq}\eeq
and $d=4$ for our purposes.
On the small-$\beta$ side of the transition, one expects
the strong coupling series for $\P$ to work; this is an expansion of $\P$
in powers of $\beta^2$ \cite{drouffe}. At larger $\beta$ one expects
renormalization group running of $\P$ \cite{lepage}. For $N_c=4$ the
change from strong to weak coupling behaviour is fairly smooth, with a
cross over in the vicinity of $\beta=10.2$ (see Figure \ref{fg.bulkco4}).
The strong coupling side has little to do with continuum physics. We study
thermal physics on the weak-coupling side of this crossover, where,
as we show in Section \ref{sc.rgscl}, the continuum limit can be taken.

The largest finite volume effect at $T=0$ is expected to occur when the
lattice sizes are such that a spurious deconfinement transition takes
place \cite{saumen}. At any given bare coupling $\beta$, there is a
critical $N_*(\beta)$ such that for $N^4$ lattices with $N>N_*(\beta)$
one expects small finite size effects. These small effects are expected to
scale as $\exp(-\ell m_0)$ where $m_0$ is the mass of the lowest glueball.
For SU(3) pure gauge theory, this mass is very high compared to the
deconfinement temperature $T_c$. If this happens also for $N_c>3$, then
one expects that finite size effects should be smaller than of order
$\exp[-N/N_*(\beta)]$. That finite volume effects are indeed small at $T=0$
is borne out by the data in Table \ref{tb.pl46}.

The finite temperature transition was monitored using the order parameter
Polyakov loop, $\L$, where
\beq
  L = \frac1{N_s^3}\sum_i \tr\prod_{t=1}^{N_t} U_{i,\hat t},
\label{poloop}\eeq
where the sum over sites, $i$, is restricted to all spatial sites.
The order parameter jumps from a zero value at small temperature to a
finite value at the thermal transition, signaling deconfinement. The
thermal transition is of first order in all the simulations presented
here. We found that for SU(4) and SU(6) gauge theories the finite
temperature transition and the bulk transition interfere for $N_t=4$
(see Figure \ref{fg.bulk}). This is a known phenomenon \cite{nc4,datta}.
Since the
bulk transition must occur at a fixed lattice spacing, it is natural to
expect that by changing $N_t$ the bulk and the thermal transitions can
be decoupled. It was found \cite{nclarger} that at larger $N_t$ these transitions do
separate out (see Figure \ref{fg.betac4}). Therefore, our strategy in
this paper is to study larger $N_t$, where the thermal transition is in the
weak coupling regime, and use these studies to take the continuum limit.

\section{The deconfinement transition}
\label{sc.deconf}

\begin{figure}
\begin{center}
   \scalebox{0.7}{\includegraphics{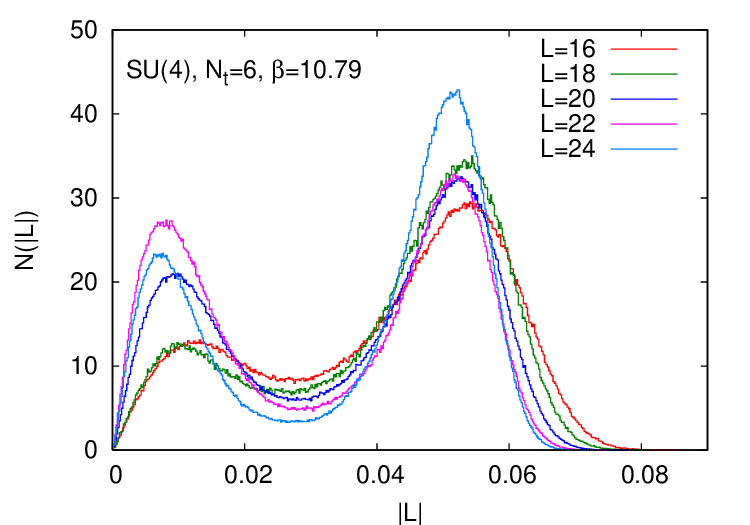}}
   \scalebox{0.7}{\includegraphics{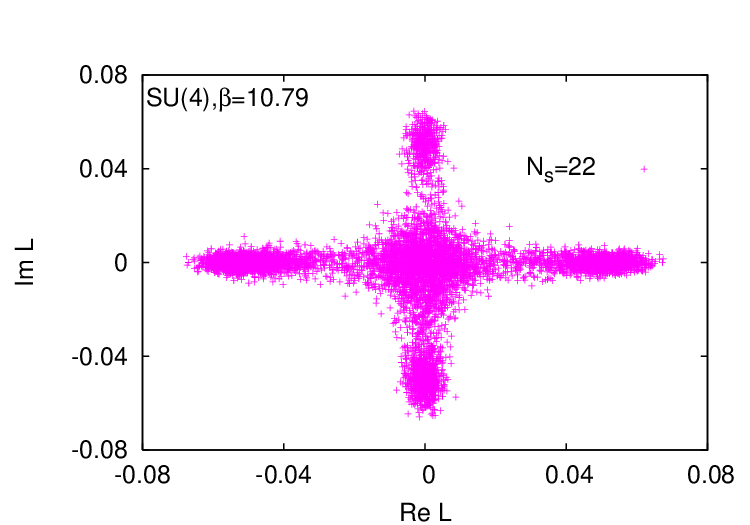}}
\end{center}
\caption{A first order transition, \ie, the coexistence of phases, with
 different values of $\L$, is signaled by a multi-peaked
 histogram of $|L|$ and the fact that the scatter plot of $L$ in the
 complex plane shows 5 well developed dense regions--- $L=0$ and four
 complex values of $L$. Here we show these features at a coupling where
 all five coexisting phases for the SU(4) theory have large weight.}
\label{fg.metastab}
\end{figure}

\begin{figure}
\begin{center}
   \scalebox{0.7}{\includegraphics{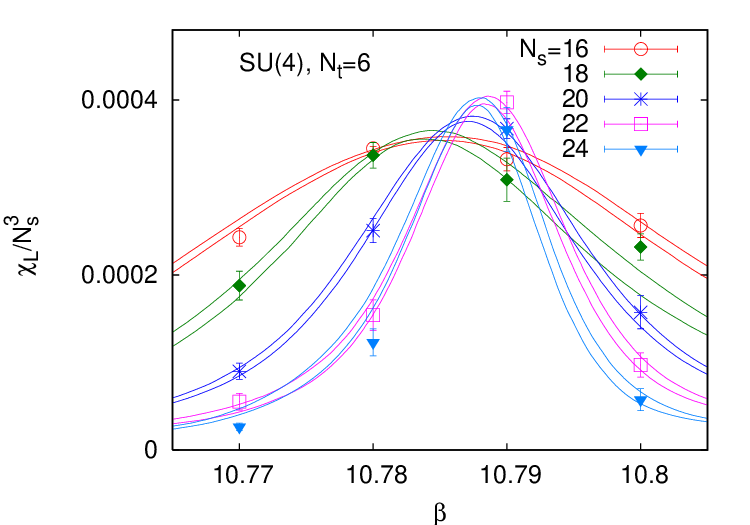}}
   \scalebox{0.7}{\includegraphics{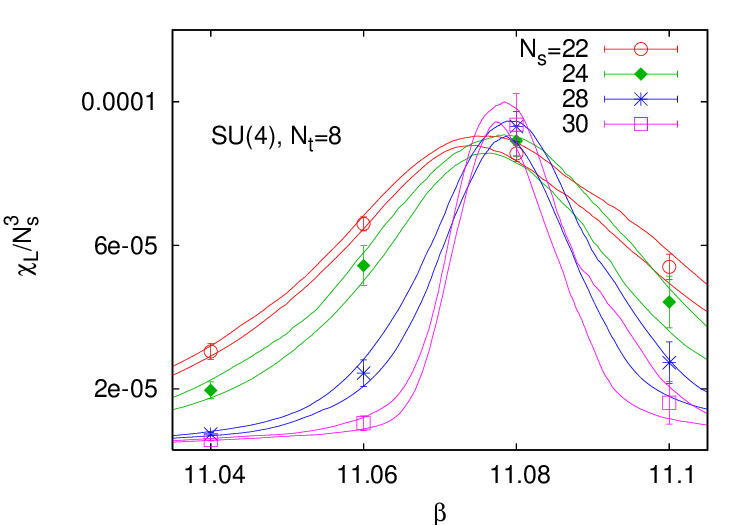}}
\end{center}
\caption{Re-weighting analyses of $\chil$ for SU(4) gauge theory shows that
  on the larger lattices the maximum scales with the lattice volume, $N_s^3$,
  indicating a first order phase transition. The analysis is shown for both
  $N_t=6$ and 8.}
\label{fg.su4rewt}
\end{figure}

\begin{figure}
\begin{center}
   \scalebox{0.6}{\includegraphics{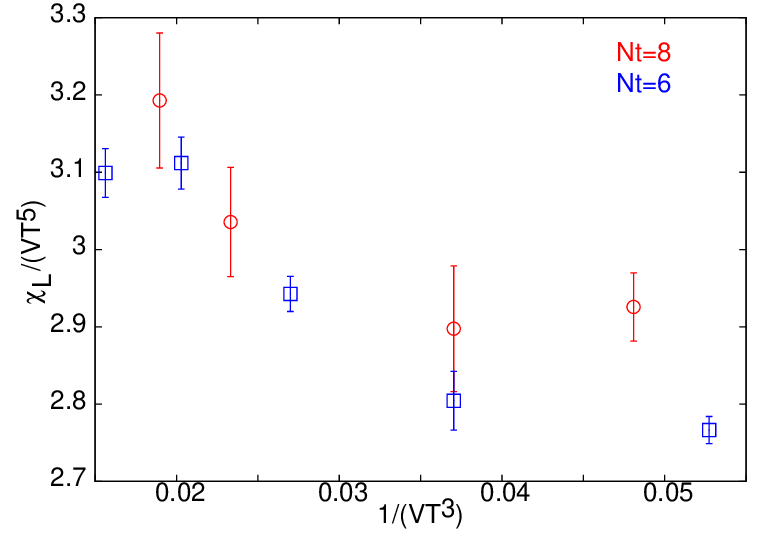}}
   \scalebox{0.6}{\includegraphics{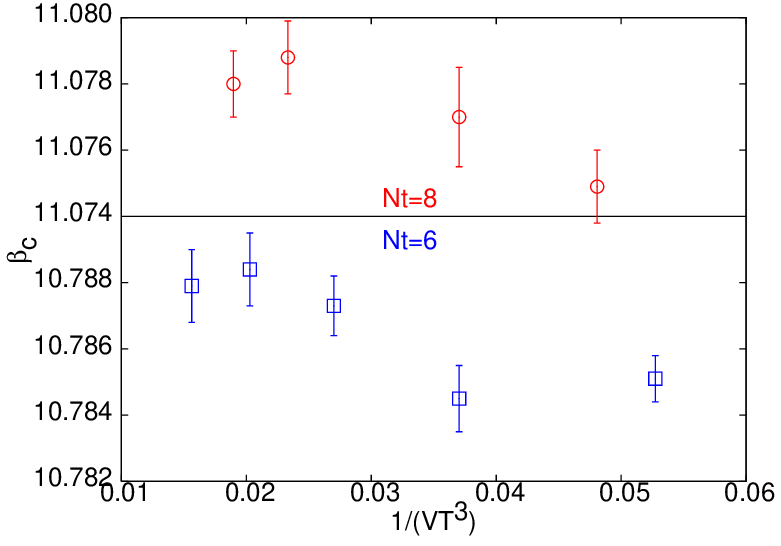}}
\end{center}
\caption{Finite size scaling for SU(4) gauge theory for $N_t=6$ (boxes) and
  8 (circles). The first panel shows the maximum of $\chil/(VT^5)$ as a
  function of $1/(VT^3)$. The second panel shows $\beta_c$ as a function
  of $1/(VT^3)$. On the largest spatial volumes, the maximum scales as $V$,
  as expected for a first order phase transition. On the same volumes
  $\beta_c$ reaches a limit which is its thermodynamic value.}
\label{fg.fsesu4}
\end{figure}

\begin{figure}
\begin{center}
   \scalebox{0.7}{\includegraphics{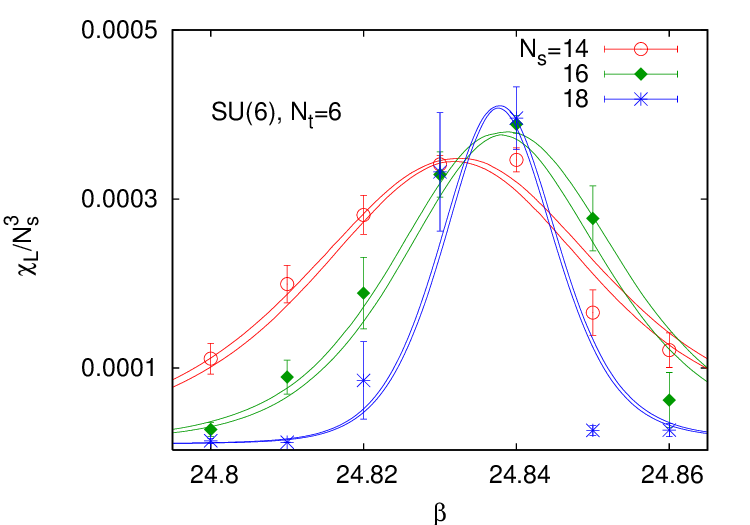}}
   \scalebox{0.7}{\includegraphics{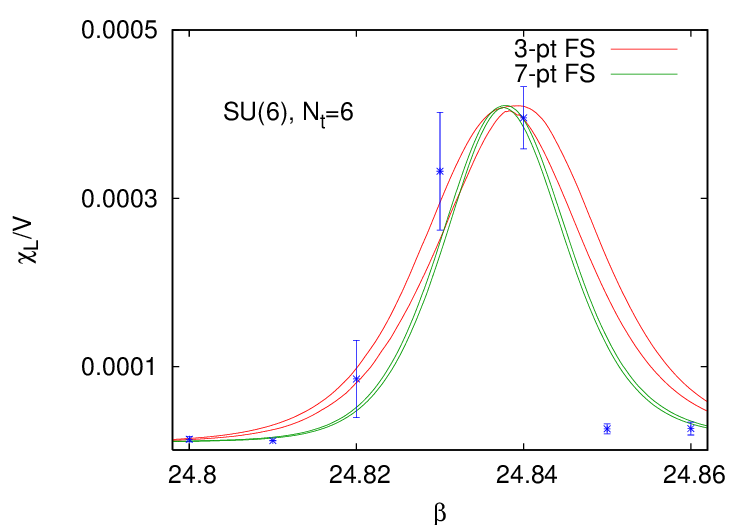}}
\end{center}
\caption{Re-weighting analyses of $\chil$ for SU(6) gauge theory shows that
  on the larger lattices the maximum scales with the lattice volume, $V$,
  indicating a first order phase transition. Also shown is a multi-histogram
  analysis on the largest lattice with seven and three input simulations,
  demonstrating the stability of the estimate of $\beta_c$.}
\label{fg.su6rewt}
\end{figure}

The abrupt change of $\L$ shown in Figure \ref{fg.betac4} indicates that
the finite temperature transition could be of first order.
Clear evidence of the coexistence of phases labeled
by the value of $\langle L\rangle$ is obtained from the distribution of $L$.
In simulations of the SU(4) theory close to
$\beta_c$ we found that the system is equally likely to be in the phase
with $L=0$ and in four phases with the same $\langle|L|\rangle$ but
different phase angles (Figure \ref{fg.metastab}). Hence the histogram
of $|L|$ shows two peaks, one close to zero and another elsewhere.
A scatter plot of $L$ measured on
each gauge field configuration also shows four distinct populations. All
these observations are consistent with a first order phase transition.
The extraction of the jump in $\langle L\rangle$ at $T_c$ needs the
renormalized Polyakov loop \cite{kay} and hence lies beyond the scope of
this study.

For more accurate determination of $\beta_c$ we defined this coupling
by the position of the maximum of the susceptibility of $|L|$---
\beq
   \chil = N_s^3 \left\{
     \left\langle|L|^2\right\rangle-\left\langle|L|^2\right\rangle\right\}.
\label{chil}\eeq
For the exploratory runs marked in Table \ref{tb.runs},
$\beta_c$ is estimated from the position of the peak of
the values of $\chil$ found in a scan over $\beta$,  and its quoted
error is the spacing
in the scan of $\beta$. In all the remaining cases, the objective was
precision, and maximum of $\chil$ was determined through multi-histogram
re-weighting
\cite{ferrenberg}. The errors on $\chil$ were defined through a bootstrap
procedure combined with the re-weighting. Such analysis requires very
large statistics, which is available to us, as shown in Table
\ref{tb.rewt}.

A final verification of the order of the transition and the determination
of the critical coupling, $\beta_c$, require finite size scaling \cite{fsep}.
At a first order transition the maximum value of $\chil$ as a function of
$\beta$ should scale as $N_s^3$, \ie, 
\beq
   \chil^m(N_s^3) = \alpha N_s^3 + \gamma + {\cal O}(N_s^{-3}),
\label{chilmscale}\eeq
when $N_s^3$ is large enough. Also, the position of the peak, which is our
estimate of $\beta_c$ at finite volume, should scale as
\beq
   \beta_c(N_s^3) = \beta_c + \delta N_s^{-3} + {\cal O}(N_s^{-6}),
\label{betacscale}\eeq
as one approaches the thermodynamic limit, $N_s\to\infty$. A different
definition of $\beta_c(N_s^3)$, such as the one where the $N_c+1$ different
peaks in $L$ have equal weight, could give a different
result at finite $N_s^3$ through a change in $\delta$. 
Finite volume scalings as in eqs.\ (\ref{chilmscale},
\ref{betacscale}) were observed in SU(3) gauge theory \cite{nc3}. The
asymptotic region sets in when the lattice size is much larger than
the longest correlation length in the system. In this asymptotic region
one expects exponentially slow sampling through a standard Monte Carlo
procedure, $\tau_{int}\propto\exp(\sigma V^{2/3})$ \cite{tauint}.
As a result, one might expect that as the transition becomes stronger
it becomes harder to do a finite size scaling analysis because of an
increase in $\sigma$, but the asymptotic finite volume corrections,
$\gamma$ and $\delta$, also become relatively smaller.

The variation of $\chil$ with $\beta$ obtained through a bootstrap
multi-histogram analysis is shown for the SU(4) theory with $N_t=6$ and
8 in Figure \ref{fg.su4rewt}. 
The position of the peak of $\chil$,
\ie, $\beta_c$, is very stable on the largest lattices used,
as shown in Figure \ref{fg.fsesu4}. 
It seems that on the two
or three largest lattices one
enters the region of asymptotic finite size scaling where the formulae in
eqs.\ (\ref{chilmscale}, \ref{betacscale}) become applicable. 
The values of
$\beta_c$ for $N_t=6$ and 8, shown in Table \ref{tb.betac}, are obtained
by fitting eq.\ (\ref{betacscale}) with the constraint $\delta=0$
to data on the three largest volumes at each $N_t$. If $\delta$ is
allowed to vary freely then the best fit changes by at most the quoted
error and we find $\delta=-14\pm14$ for $N_t=6$ and $\delta=-32\pm38$
for $N_t=8$.\footnote{The estimates of the thermodynamic limits of
$\beta_c$, extrapolated from smaller lattices in \cite{nclarger}, are
in rough agreement with ours.  Their results for $\delta$ are 
significantly different from zero but compatible
with the values we get including our smaller lattices. Such a
volume dependence of $\delta$ indicates that the leading terms in eq.\
(\ref{betacscale}) are not sufficient to parametrize the shifts over this
large a range in $\zeta^3$. The rough agreement of the thermodynamic limit
of $\beta_c$ in the two cases can then be attributed to an overall small
finite volume shift, as may happen for a strong first-order transition.}

For the SU(4) theory with $N_t=10$ and 12 we have performed simulations
on only one lattice volume, as shown in Table \ref{tb.runs}. While the
multi-histogram reweighting analysis allows us to find $\beta_c$ at this
volume with good precision, an extrapolation to the thermodynamic limit
is not yet possible. If we were to assume that $\delta=0$, as in the
two smaller $N_t$ sets we discussed above, then we can ignore the finite
volume shift. However, for $N_t=10$ and 12 we use $\zeta$ smaller than
the lattices which gave $\delta=0$ for $N_t=6$ and 8, so there may be some
finite volume shift. To estimate this, albeit crudely, we fitted eq.\
(\ref{betacscale}) to our data on the three smallest volumes for $N_t=6$
and 8, and extrapolated $\delta$ to $N_t=10$ and 12 using a scaling
formula for $\delta$ in \cite{nclarger}. According to this analysis, the
thermodynamic limit of $\beta_c$ is within twice the error quoted for
$N_t=10$ and within the quoted errors for $N_t=12$.

SU(6) follows the same trend. For all $N_t$, one has all the qualitative
features of a strong first order phase transition--- multiple coexisting
phases (6 ordered phases and one disordered in this case) and long
auto-correlation times determined by the tunneling rate from one phase
to another, growing rapidly with volume. The phase transition is even
stronger than SU(4), and a finite size scaling analysis is more delicate.

In Figure \ref{fg.su6rewt} we show the multi-histogram reweighting
analysis for SU(6). Note that the aspect ratios used in this analysis are
smaller than those for SU(4). This is forced on us because the transition
is stronger, and therefore $\tau_{int}$ grows faster with $V$. In fact,
statistical problems already begin to show up at the largest $V$
for $N_t=6$; the run at $\beta=24.85$ has statistically too few tunnelings,
since it lies right at the edge of the region of metastability for these
lattices. For this system we examined the stability of the analysis
through the comparison of the multi-histogram method with seven and
three histograms.  As shown in Figure \ref{fg.su6rewt}, the peak is
unambiguously determined, since the values of
$\chil^m$ in the two analyses are compatible, as are the estimates of
$\beta_c$.  The reason for the absence of a large systematic error at the
peak is that the scan in $\beta$ is fine enough, so that there are enough
other histograms to compensate for the one which is improperly sampled.

\begin{table}
\begin{center}
\begin{tabular}{|c|lllll|}
\hline
$N_c$ & $N_t=4$ & $N_t=6$ & $N_t=8$ & $N_t=10$ & $N_t=12$ \\
\hline
3 & 5.6925 (2) & 5.8940 (5) & 6.0609 (9) & & \\
4 && 10.788 (1) & 11.078 (1) & 11.339 (4) & 11.552 (17) \\
6 && 24.838 (1) & 25.470 (3) & 26.0 ($1^*$) & \\
8 && 44.7 ($2^*$) & 45.8 ($2^*$)  & & \\
10 && 70.5 ($15^*$) & 73 ($2^*$)  & & \\
\hline
\end{tabular}
\end{center}
\caption{The critical couplings, $\beta_c$, for the first order thermal
phase transition for different $N_c$ and different temporal lattice
sizes, $N_t$. Error estimates which are marked by an asterisk are not
statistical, as discussed in the text. The results for $N_c=3$ were
found in \protect\cite{nc3}. For $N_c>3$ the transition for $N_t=4$ falls
in the region of the strong to weak coupling cross over, making it hard
to distinguish the bulk from the thermal phase transition.}
\label{tb.betac}
\end{table}

\begin{figure}
\begin{center}
   \scalebox{0.7}{\includegraphics{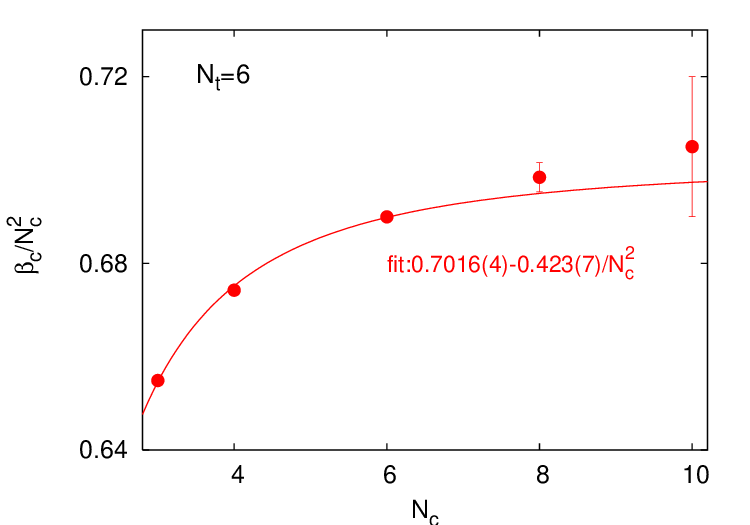}}
   \scalebox{0.7}{\includegraphics{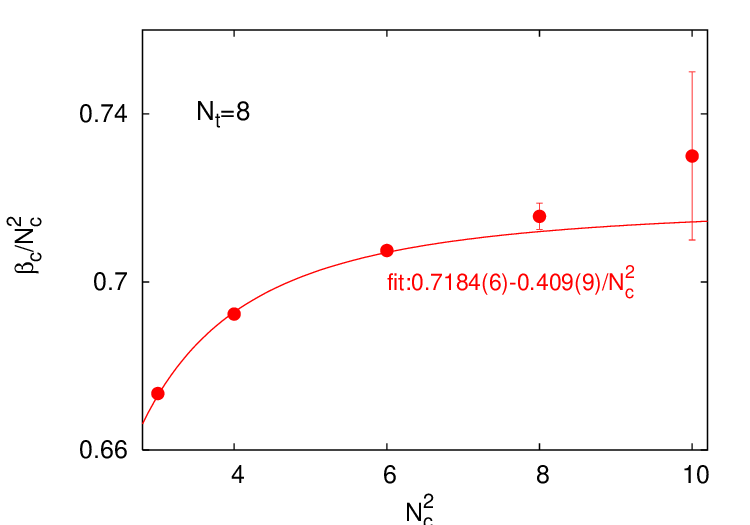}}
\end{center}
\caption{$\beta_c$ for different number of colors, on $N_t$ = 6 and 8 lattices.}
\label{fg.ncscaling}
\end{figure}

Our simulations of SU(8) and SU(10) pure gauge theory at finite
temperature were purely exploratory, being restricted to a single volume at
each $N_t$. The value of $\beta_c$ that we estimate, along with the error
bounds given by the scan in $\beta$ are quoted in Table \ref{tb.betac}.

In Figure \ref{fg.ncscaling} we plot these results as a function for $N_c$
at fixed lattice spacing $a=1/(N_t T_c)$ for $N_t=6$ and 8. We see that a
good description of our observations is obtained by a two-term extrapolation
to the large-$N_c$ limit---
\beq
   \frac{\beta_c}{N_c^2} = \beta_* + \frac{\beta'_*}{N_c^2}.
\label{extrap}\eeq
The quantity $\beta_*$ is expected to increase without bound as $N_c\to\infty$.
The first correction term, of order $1/N_c^2$,
provides a sufficient description of the data even at $N_c=3$.
This scaling check shows that for each cutoff, $a=1/(N_tT)$ one has a large
$N_c$ theory which is non-trivial in the limit $g^2N_c$ fixed, \ie,
$\beta/N_c^2$ fixed.

Note that in the best cases we have achieved accuracies of a few parts in
$10^4$ in the measurement of $\beta_c$. Next we turn to the continuum
extrapolation of these measurements and the determination of the temperature
scale.

\section{Renormalized coupling and the temperature scale}
\label{sc.rgscl}

Pure gauge SU($N_c$) theory contains a single dimensionless parameter, the
coupling, $\alphas=g^2/4\pi$.
Quantum corrections change this into a scale. This can be specified
explicitly, as the parameter $\Lambda$, or implicitly, as the value
of the running (renormalized) coupling $\alphas(\mu)$ at a chosen
momentum scale $\mu$.
At scales where $\alphas$ is small, perturbation theory is expected to work.
In that case, changes of the scale of measurements can be accomplished
through the use of
perturbation theory. In particular, extrapolation of results to the continuum
can then be done with ease.

The two-loop RGE can be integrated to trade the running coupling
$\alphas(\mu)$ for a mass scale, 
\beq
   a\Lambda = k R\left(\frac1{4\pi\beta_0\alphas}\right),
      \qquad{\rm where}\qquad
   R(x)=\exp(-x/2)x^{\beta_1/(2\beta_0^2)},
\label{scale}\eeq
where $k$ depends on the coupling $\alphas$ that enters into these equations.
This coupling is measured by some operator dominated by the ultraviolet scale
$1/a$.  Each such definition of $\alphas$ defines an RG scheme. The function $R$
is obtained by integrating the two-loop beta function, 
\beq
\overline\beta(g) = \mu \frac{dg}{d\mu} = -\beta_0g^3-\beta_1g^5,
\label{betafn}\eeq
where $\beta_0$ and $\beta_1$ are well-known \cite{pdg}. These coefficients
are independent of the scheme. Since $T=1/(aN_t)$, and we have a determination
of $T_c$ for different $N_t$, by making appropriate lattice measurements of
$\alphas$ we can measure the temperature scale, $T/T_c$. At the same time,
one could use eq.\ (\ref{scale}) to determine the QCD scale $\lms$ in terms
of $T_c$.

In order to complete this process, we need to define $\alphas$. Two schemes
are easily implemented on the lattice. One is the $V$ scheme \cite{lepage},
in which the potential extracted from Polyakov loop correlations is used to
define the renormalized coupling. Equivalently, at two-loop order accuracy,
the weak-coupling expansion of the plaquette \cite{klassen} can be
inverted to find $\alphav$---
\beq
  -\ln\P = \pi C_F\alphav(q)\left[1 -\frac{11N_c}{12\pi} 
     \ln\left(\frac{6.7117}{aq}\right)^2 \alphav(q) \right]
\label{plaqv}\eeq
where $C_F=(N_c^2-1)/(2N_c)$ and $q=k/a$, where $k$ is the same number which
is used in eq.\ (\ref{scale}). In this scheme $k=3.4018$ \cite{lepage}. Since
$\P$ is easily measured and needed for thermodynamic quantities, we prefer
to use eq.\ (\ref{plaqv}) as a definition of $\alphav$ rather than through a
separate measurement of the potential. The
other definition is the E-scheme, in which the coupling is defined from the
plaquette through the formula 
\beq
  1-\P=\pi C_F\alphae(q),
\label{escheme} \eeq
where $q=1/a$, \ie, $k=1$.
If the weak coupling expansion were exact, and known to all orders, then
there would be no difference between the couplings determined in these
two schemes at any cutoff, provided that $\alphav$ (or $\alphae$) were
small enough. Since this is not the case, one must explore RG scheme
dependence. A third scheme that we utilize is the $\overline{\rm MS}$
scheme defined through dimensional regularization of the continuum
perturbation theory. The known expansion of $\alphav$ in terms of
$\alphams$ \cite{peter} is used to obtain the latter using the two-loop
relation
\beq
   \alphams(q')=\alphav(q)\left[1+\frac{2N_c}{3\pi}\alphav\right],
\label{msbarscheme}\eeq
where $q'=\exp(-5/6)q$ \cite{brodsky}. In other words, $k=1.4784$ for
the $\overline{\rm MS}$ scheme.

The values of the plaquette at zero temperature are measured on the grid
of $\beta$ shown in Tables \ref{tb.pl46} and \ref{tb.pl810}. They are
obtained at other points using Lagrange interpolation with polynomials
of orders between 1 and 4, and through a cubic spline interpolation. By
using such a variety of interpolation schemes we quantify the systematic error
in the interpolation at any $\beta$ as the widest dispersion between these
schemes.  For SU(4) and SU(6) on lattices with $N_t\ge6$, this systematic
error is smaller than, or of the same order as, the statistical error in
the measurement of the plaquette.  For SU(8) and SU(10), the systematic
error is larger than the statistical error. These lead to statistical and
systematic errors in the determination of the running coupling of the
order of a few parts in $10^5$. However, when we determine a scale,
the largest error is that which comes from the determination of $\beta_c$.

A test of the weak coupling expansion for the scale, and the scheme
dependence in this is provided by using the determination of $\beta_c$
for one $N_t$ to predict that at a different $N_t$. Since we have
measurements for $N_t=6$, 8 and 10 for SU(4) and SU(6), we have chosen to
examine the temperature predicted by the one-loop and two-loop RGEs for
the $N_t=6$ and 10 lattices at the $\beta_c$ corresponding to the $N_t=8$
lattice. This is shown in Table \ref{tb.rgsdep}. Note that the error of roughly
one part in $10^4$ in the determination of $\beta_c$ translates into an
error of about one part in $10^3$ in the determination of the temperature
scale in the range of temperatures we explore here.
Since the accuracy of this error estimate is important in our later
reasoning, we performed it by two different methods: first by the usual
methods of propagating errors, and then again through a bootstrap
analysis. The two errors agreed within 10\%, indicating that the estimates
are robust. The errors quoted in Table \ref{tb.rgsdep} come from
the bootstrap analysis.

\begin{table}[bth]
\begin{center} 
\begin{tabular}{|c|c|c|c|c|}
\hline
$N_c$ & $N_t$ & Scheme & 2-loop & 1-loop \\
\hline
\multirow{6}{*}{4} & \multirow{3}{*}{6} & E & 1.29709\ (167)\ (7)\ (1) & 1.32333\ (184)\ (8)\ (1) \\ 
& & V & 1.30782\ (174)\ (8)\ (1) & 1.35339\ (204)\ (9)\ (2) \\
& & $\overline{\rm MS}$ & 1.30057\ (169)\ (7)\ (1) & 1.35068\ (202)\ (9)\ (1) \\ 
\cline{2-5}
& \multirow{3}{*}{10} & E & 0.80885\ (265)\ (5)\ (0) & 0.79632\ (280)\ (5)\ (0) \\
& & V & 0.80442\ (270)\ (5)\ (0) & 0.78381\ (294)\ (5)\ (1) \\
& & $\overline{\rm MS}$ & 0.80757\ (267)\ (5)\ (1) & 0.78517\ (292)\ (5)\ (0) \\
\hline
\multirow{6}{*}{6} & \multirow{3}{*}{6} & E & 1.30504\ (165)\ (10)\ (3) & 1.33190\ (181)\ (11)\ (4) \\
& & V & 1.31675\ (172)\ (10)\ (4) & 1.36476\ (200)\ (12)\ (4) \\
& & $\overline{\rm MS}$ & 1.30916\ (167)\ (10)\ (3) & 1.36130\ (198)\ (12)\ (4) \\
\cline{2-5}
& \multirow{3}{*}{10} & E & 0.81884\ (2993)\ (4)\ (1) & 0.80695\ (3161)\ (4)\ (1) \\
& & V & 0.81435\ (3052)\ (4)\ (1) & 0.79452\ (3324)\ (4)\ (1) \\
& & $\overline{\rm MS}$ & 0.81740\ (3011)\ (4)\ (1) & 0.79583\ (3307)\ (4)\ (1) \\
\hline
\end{tabular}
\end{center}
\caption{The values of $T/T_c$ at the coupling $\beta_c(N_t=8)$, for
  SU(\/$N_c$\/) gauge theory for different $N_t$, in different RG schemes,
  and at different loop orders. The entries give the central value, the
  statistical error propagated from the uncertainty in $\beta_c(N_t)$, the
  statistical error from plaquette measurements, and systematic errors from
  interpolations of plaquette values. For $N_t=6$ the exact non-perturbative
  result is $T/T_c=1.33$, and for $N_t=10$ it is $T/T_c=0.80$.}
\label{tb.rgsdep}\end{table}

Some systematics visible in Table \ref{tb.rgsdep} is worth explicit
comment. The one-loop RG already is close to the exact result, but in
all cases performs worse than the two-loop RG. This is expected. Also the RG
flow between $\beta_c(N_t=10)$ and $\beta_c(N_t=8)$ is better than that
between $\beta_c(N_t=6)$ and $\beta_c(N_t=8)$. Indeed, for SU(4), where the
test is most stringent, the former agrees with the exact non-perturbative
result to about $1.5\sigma$ in the V-scheme, and to about 3--4$\sigma$
in the other schemes. The implication is that $N_t=8$ is already in a
regime where the weak coupling extrapolation to the continuum works, but
$N_t=6$ may be just a little outside this regime. The scheme dependence
is most significant when the two-loop RGE is least reliable, but is less
than 1\% in all cases.

\begin{figure}
\begin{center}
   \scalebox{0.7}{\includegraphics{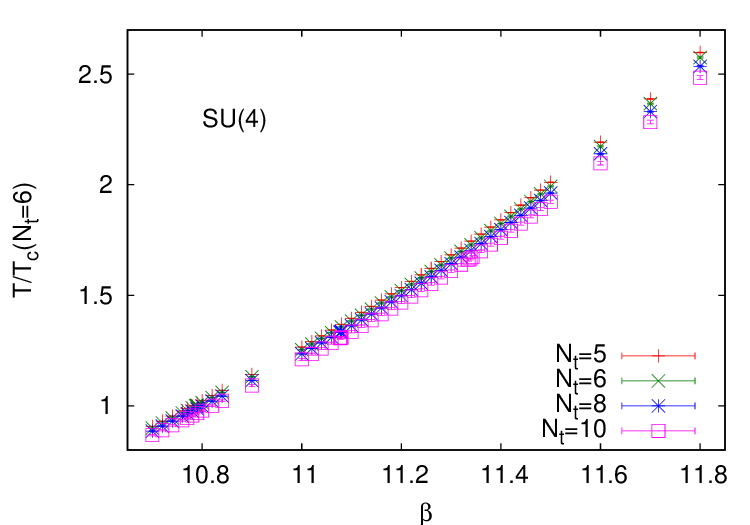}}
   \scalebox{0.7}{\includegraphics{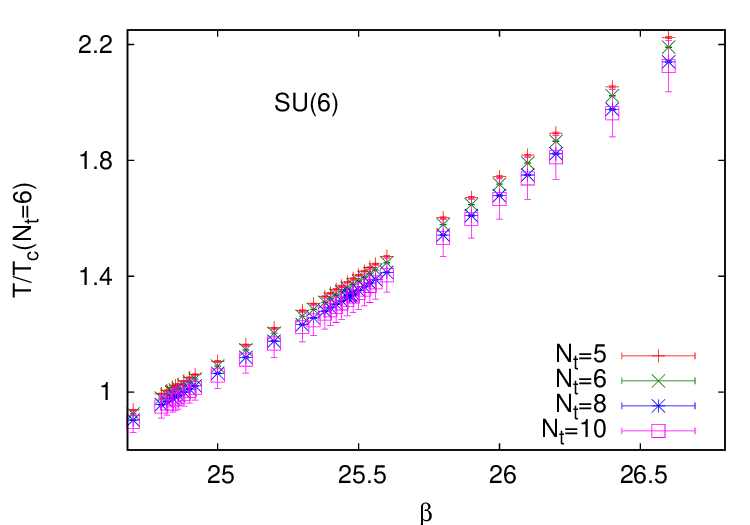}}
\end{center}
\caption{One-loop renormalization group flow for $N_c=4$ and
 6 in the V-scheme. The data on $\beta_c$ for $N_t=5$ is taken
 from \protect\cite{nclarger}. If the RG were adequate, then the
 curves for different $N_t$ would lie on top of each other. The
 accuracy of the one-loop flow improves with increasing $N_t$.}
\label{fg.1rgflow}
\end{figure}

\begin{figure}
\begin{center}
   \scalebox{0.7}{\includegraphics{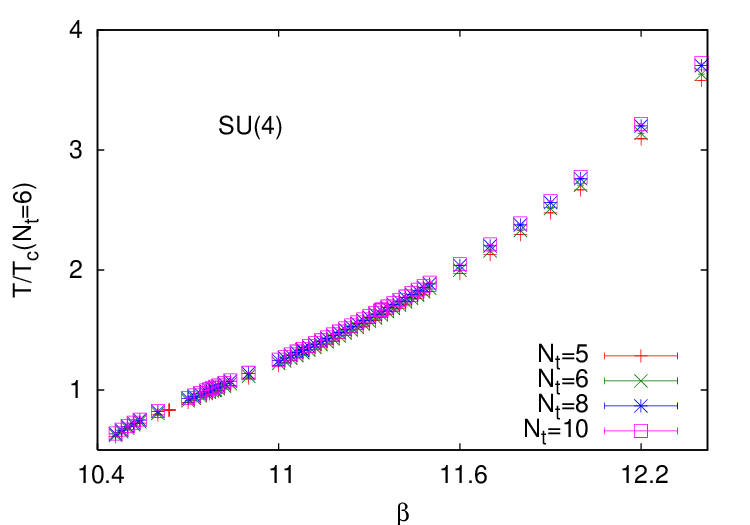}}
   \scalebox{0.7}{\includegraphics{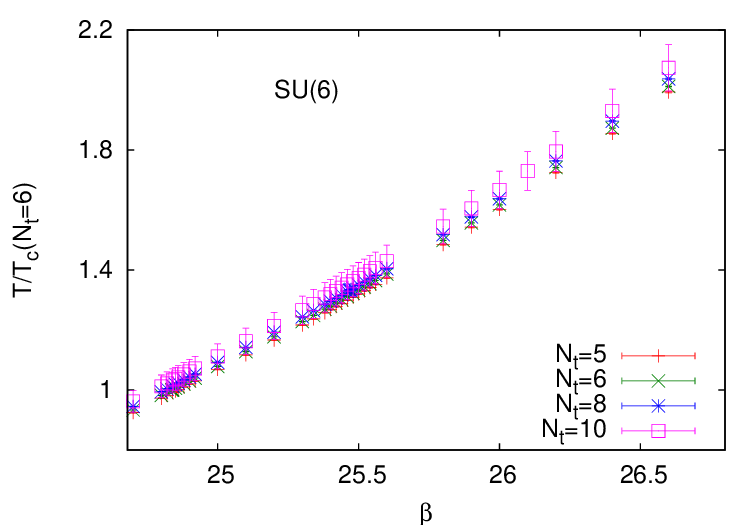}}
\end{center}
\caption{Two-loop renormalization group flow for $N_c=4$ and
 6 in the V-scheme. The data on $\beta_c$ for $N_t=5$ is taken
 from \protect\cite{nclarger}. Good scaling behaviour is obtained when
 the curves for different $N_t$ lie on top of each other.
  Excellent scaling is obtained for $N_t=8$ and 10.}
\label{fg.2rgflow}
\end{figure}

The one-loop temperature scale is shown in Figure \ref{fg.1rgflow} in
the V-scheme for a large range of lattice spacings.  While this works
reasonably well, the improvement in going to two-loops, shown in Figure
\ref{fg.2rgflow}, is obvious.  The cutoff effects are small on this scale
already for $N_t=5$.  The two-loop temperature scale in the V-scheme
for $N_t=6$ and 8, are collected together in Table \ref{tb.tbtc} for
future reference. As discussed already, the scale for $N_t=8$ is more
reliable, and should be used for extrapolations. The scale for $N_t=6$
serves to give a rough indication of the kind of systematic errors to be
expected: as one can see the difference between these two scales is roughly
of 2\%.

We have argued here that the continuum extrapolation from $N_t=8$ or 10 can
be performed using the weak coupling expansion. In weak coupling
the reference scale that is used is $\lms$ and not $T_c$. Our analysis above
gave us the scales $\Lambda_E$ and $\Lambda_V$. These can be converted into
$\lms$ \cite{alles,luscher}, 
\beq
  \frac{\lms}{\Lambda_E} = \exp\left[\frac{6\pi}{11}
       \left(1.622268-\frac\pi{4N_c^2}\right)\right],\qquad\qquad
  \frac{\lms}{\Lambda_V} = \exp\left[-\frac{31}{66}\right].
\label{lle}\eeq
Using this and the determinations of $\beta_c$, we can convert the
non-perturbative scale $T_c$ into a specification of $\lms$ in three
different schemes.

\begin{table}
\begin{center}
\begin{tabular}{|c|cc|cc|cc|}
\hline
$N_c$ & \multicolumn{2}{|c|}{E-scheme}
      & \multicolumn{2}{|c|}{V-scheme}
      & \multicolumn{2}{|c|}{$\overline{\rm MS}$-scheme} \\
 & $T_c/\lms$ & $c_2$ & $T_c/\lms$ & $c_2$ & $T_c/\lms$ & $c_2$ \\
\hline
3 & 1.16 (2) & 2.7 (7) & 1.11 (2) & 1.9 (7) & 1.17 (2) & 2.5 (7) \\
4 & 1.198 (1) & 2.49 (5) & 1.129 (2) & 1.61 (5) & 1.203 (2) & 2.23 (5) \\
6 & 1.193 (1) & 1.98 (9) & 1.120 (2) & 1.05 (9) & 1.193 (2) & 1.67 (9) \\
\hline
\end{tabular}
\end{center}
\caption{Fitted parameters for the non-perturbative beta function in the
 form of eq.\ (\ref{fit}). Here $T_c/\lms$ must be considered as a formal
 fit parameter. Data from all available lattice spacings
 $1/(12T_c)\le a\le1/(6T_c)$ have been used. For $N_c=3$ data from
 $a=1/(4T_c)$ has also been used.}
\label{tb.ehk}
\end{table}

\begin{table}
\begin{center}
\begin{tabular}{|c|ccc|}
\hline
$N_c$  & E-scheme & V-scheme & $\overline{\rm MS}$-scheme \\
\hline 
3 & 1.19 (3) & 1.12 (3) & 1.20 (2) \\
4 & 1.235(1) & 1.153(1) & 1.236(1) \\
6 & 1.222(1) & 1.135(1) & 1.217(1) \\
8 & 1.26(6) & 1.17(5) & 1.25(6) \\
10 & 1.48(41) & 1.38(39) & 1.48(41) \\
\hline
$\infty$ & 1.22 & 1.13 & 1.22 \\
\hline
\end{tabular}
\end{center}
\caption{$T_c/\lms$ in the continuum limit of SU($N_c$) gauge theory for
$N_c$ = 3,4,6,8, in different schemes using two-loop RGE for $a\le1/(8T_c)$.
These values of $T_c/\lms$ are appropriate for use in a two-loop computation.}
\label{tb.tclms}
\end{table}

The test of two-loop RGE in Table \ref{tb.rgsdep} showed that this
was fairly accurate already at the lattice spacing corresponding to
$\beta_c$ for $N_t=8$ and 10. Therefore it is no surprise that to the
same degree of accuracy the ratio $T_c/\lms$ is constant when evaluated
at $N_t\ge8$.  However, the scheme dependence is much larger for $\lms$
than for the temperature scale. This happens because the renormalized
coupling is not small enough for the eqns.\ (\ref{lle}) to hold. One
could correct these formulae by explicitly including two-loop or higher
order correction terms.  However, then the ratio $T_c/\lms$ would depend
on the scale $a$. This can be avoided only when $\alphas$ becomes
substantially smaller. However, since $\alphas$ runs logarithmically
with $a$, that would imply that one has to use lattice spacings which
are about 10 times smaller. This is currently outside the reach of our
computational abilities.

For SU(3) the range of accuracy of the RGE can be extended by including
into it corrections of order $a^2$ \cite{ehk}. This seems to be possible
for $N_c>3$ too. When we compare $T_c/\lms$ extracted for all $N_t$, it
seems possible to fit this to a simple $a^2$ variation. In principle
this term can be used to add an ${\cal O}(a^2)$ correction to the two-loop
beta function by changing $R(x)$ in eq.\ (\ref{scale}) to $R(x)[1+\eta/N_t^2]$.
We evaluate these corrections by a fit to the lattice spacing dependence of
$T_c/\lms$ which renders this ratio flat in the whole range $1/(10T_c)\le a
\le1/(5T_c)$, \ie, we choose the fit form
\beq
 \frac{T_c}{\lms}|_{N_t} = \frac{T_c}{\lms} + \frac{\eta}{N_t^2}.
\label{fit}\eeq
Statistically significant results can only be obtained for $N_c\le6$. Our
results for the fit are given in Table \ref{tb.ehk}. One can compare
these with the estimate $T_c/\lms=1.187\pm0.009$ obtained by combining
estimates of $T_c/\sqrt\sigma$ (where $\sigma$ is the string tension) and
$\sqrt\sigma/\lms$ reported in \cite{nclarger,allton}. The estimation of
$\sqrt\sigma/\lms$ removes ${\cal O}(a^2)$ corrections, as we do here. We
note that such a term sums many different types of corrections and
amounts to a phenomenological fit of the beta function, \ie, gives
what is called the non-perturbative beta-function. For this reason
it cannot be regarded as a test of scaling.

The continuum values for $T_c/\lms$, obtained assuming that this ratio
is constant for lattice cutoffs $a\le1/(8T_c)$ are collected in Table
\ref{tb.tclms}. Note that there are large and statistically significant
differences between these results and those in Table \ref{tb.ehk}. Since
the latter results constitute a check of two-loop RGE, and the best possible
extraction of $\lms$, they are to be preferred for this purpose. For
SU(3) we have performed a re-analysis of the data which was used in
\cite{scaling} without the ${\cal O}(a^2)$ terms from \cite{ehk}. This
makes the analysis uniform for all $N_c$. Note that the dependence
on $N_c$ is weak. We have added indicative values of this ratio
extrapolated to the limit $N_c\to\infty$. Since a statistical analysis
is not possible, we have not added error bars to this extrapolation. Note
that the strong scheme dependence, which we discussed before, propagates
to the $N_c\to\infty$ limit.

In summary, the two-loop renormalization group equations work well for
$a\le1/(8T_c)$, \ie, at the level of 1--2\%.  Since the largest part
of this uncertainty stems from the RG scheme dependence, higher order
corrections in the perturbation series for the plaquette could easily
improve this description. However, trading the non-perturbative scale
$T_c$ for the perturbatively determined scale $\lms$ is not yet possible
to better than 5--10\%.  Improving this would require using lattice
spacings which are beyond reach today.

\section{Conclusions}
\label{sc.final}

In this paper we studied the finite temperature phase transition in
SU(4), SU(6), SU(8) and SU(10) pure gauge theories at several lattice
spacings and extrapolated the results to the continuum. In all these
theories at large lattice spacing, $a\simeq1/(4T_c)$, a lattice artifact
called the bulk phase transition prevents a simple study of finite
temperature physics. The order parameter of the bulk transition is the
plaquette average, $\P$, whereas that of the finite temperature transition is
the Polyakov loop expectation value, $\L$. The bulk transition is expected
to occur at a (approximately) fixed lattice spacing. We studied these
theories at smaller lattice spacings, $a\le1/(6T_c)$, and found that in
all cases the finite temperature phase transition can be studied without
any interference from the bulk transition (see Figure \ref{fg.betac4}, for
example). More details are reported in Section \ref{sc.simul}.

We found a first order finite temperature transition for all these
theories. This was established not only
by clear signals of multiple coexisting phases labeled by different
values of $\L$, but, in several cases, also by finite size scaling tests.
These studies and also multi-histogram reweighting at fixed volumes
allowed us to locate the phase transition with precision which was
in many cases as good as a few parts in
$10^4$. Our results on the finite temperature phase transition are given
in Section \ref{sc.deconf}, and the locations of the phase transition
are collected together in Table \ref{tb.betac}.

We investigated the continuum extrapolation of our lattice results and
found that when the lattice spacing is $a\le1/(8T_c)$ then the two-loop
RGE can be used to take the continuum limit. In order to do this one has
to use a definition of the renormalized (running) coupling, called an RG
scheme. We found that when the location of the phase transition at one
lattice spacing is used to predict that at another, then the dependence
on the RG scheme is small (see Table \ref{tb.rgsdep}): the statistical
precision is about one part in $10^3$, but the scheme dependence is about
2\%. This allows us to construct a temperature scale with this degree
of precision using the non-perturbatively obtained mass scale, $T_c$.
Since the scheme dependence is the largest part of the uncertainty, higher
order corrections will reduce this error.
This is the first instance of a large $N_c$ lattice calculation
which has reached precisions good enough to test the state of the art
in the weak coupling expansion. Details of these tests can be found in
Section \ref{sc.rgscl}. One useful result is the determination of the
temperature scale in SU(4) and SU(6) gauge theories (Table \ref{tb.tbtc}).

We tried to use two-loop perturbation theory to trade the scale $T_c$
for the scale $\lms$ which is more commonly used in weak coupling
expansions, and found that the scheme dependence becomes significantly
more pronounced. The extraction of $\lms$ by this means gave statistical
errors comparable to the temperature scale, but RG scheme dependence
of about 10\%. A large scheme dependence in trading a non-perturbative
scale such as $T_c$ for the perturbative scale $\lms$ is bound to
persist in all foreseeable lattice computations.

We found that two results can be easily extrapolated to the
limit $N_c\to\infty$. The location of the critical point at fixed lattice
spacing $a=1/(N_t T_c)$ goes as $\beta_*+{\cal O}(1/N_c^2)$
for $N_c\ge3$ (see Figure \ref{fg.ncscaling}). For $T_c/\lms$ the series
could be shorter; we find no statistically significant dependence
of $T_c/\lms$ on $N_c$ in any of the three RG schemes that we studied
(see Table \ref{tb.tclms}).

These computations were carried out on the CRAY-X1 of the ILGTI in TIFR, and
on the workstation farm of the Department of Theoretical Physics, TIFR. We
would like to thank Ajay Salve for technical support.

\vfill\eject

\vfil\eject
\appendix
\section{Some details}
\label{sc.append}

Some details of the simulations and detailed tables of some of our results
are collected in this appendix.

\begin{table}[hbt!]
\begin{center}
\begin{tabular}{|r||r|r|r||r|r|r||r|r|r||r|r|r|}
\hline
 & \multicolumn{6}{c||}{$N_c=4$} &
   \multicolumn{6}{c|}{$N_c=6$} \\
\hline
 & \multicolumn{3}{c||}{$N_t=6$} &
   \multicolumn{3}{c||}{$N_t=8$} &
   \multicolumn{3}{c||}{$N_t=6$} &
   \multicolumn{3}{c|}{$N_t=8$} \\
\hline
$N_s$ & $\beta$ & Statistics & $\tau_{\rm int}$ 
      & $\beta$ & Statistics & $\tau_{\rm int}$
      & $\beta$ & Statistics & $\tau_{\rm int}$
      & $\beta$ & Statistics & $\tau_{\rm int}$ \\
\hline
   14 &&&&&& & 24.80 & 2.14 & 7354 & & & \\
      &&&&&& & 24.81 & 1.94 & 9391 & & & \\
      &&&&&& & 24.82 & 2.05 & 10538 & & & \\
      &&&&&& & 24.83 & 1.94 & 12349 & & & \\
      &&&&&& & 24.84 & 1.94 & 12113 & & & \\
      &&&&&& & 24.85 & 2.05 & 9428 & & & \\
      &&&&&& & 24.86 & 2.01 & 7747 & & & \\
\hline
   16 & 10.77 & 1.56 & 2367 & & & & 24.80 & 2.26 & 4388 & & & \\
      & 10.78 & 4.5 & 3188 & & & & 24.81 & 3.88 & 12602 & & & \\
      & 10.79 & 1.56 & 3780 & & & & 24.82 & 2.29 & 14827 & & & \\
      & 10.80 & 2.6 & 2566 & & & & 24.83 & 2.74 & 15582 & & & \\
      &&&&&& & 24.84 & 2.34 & 16984 & & & \\
      &&&&&& & 24.85 & 4.24 & 15854 & & & \\
      &&&&&& & 24.86 & 0.89 & 7907 & & & \\
\hline
   18 & 10.77 & 0.92 & 3263 & & & & 24.80 & 3.75 & 2486 & & & \\
      & 10.78 & 0.88 & 6944 & & & & 24.81 & 3.69 & 315 & & & \\
      & 10.79 & 0.88 & 5944 & & & & 24.82 & 3.62 & 15585 & & & \\
      & 10.80 & 1.8  & 4564 & & & & 24.83 & 3.68 & 18409 & & & \\
      &&&&&& & 24.84 & 4.23 & 18376 & & & \\
      &&&&&& & 24.85 & 3.62 & 3861 & & & \\
      &&&&&& & 24.86 & 2.70 & 4175 & & & \\
\hline
   20 & 10.77 & 2.8  & 2850 & & & & & & & 25.42 & 1.46 & 5533 \\
      & 10.78 & 2.9  & 6445 & & & & & & & 25.44 & 0.90 & 11482 \\
      & 10.79 & 2.8  & 8192 & & & & & & & 25.46 & 0.64 & 15773 \\
      & 10.80 & 4.4  & 5806 & & & & & & & 25.48 & 2.31 & 15951 \\
      &&&&&& & & & & 25.50 & 1.44 & 15117 \\
\hline
   22 & 10.77 & 2.1  & 3621 & 11.06 & 3.1 & 5472 &&&&&& \\
      & 10.78 & 2.1  & 6737 & 11.08 & 2.9 & 7019 &&&&&& \\
      & 10.79 & 2.1  & 10700  & & &&&&&&& \\
      & 10.80 & 3.4  & 5256  & & &&&&&&& \\
\hline
   24 & 10.77 & 0.12 & 1452 & 11.06 & 1.08 & 5959 & & & & & & \\
      & 10.78 & 2.9  & 9063 & 11.08 & 1.1 & 9821 & & & & & & \\
      & 10.79 & 2.8  & 13614  & & & & & & & & & \\
      & 10.80 & 2.8  & 4475  & & & & & & & & & \\
      &&&&&& & & & & & & \\
\hline
   28 & & & & 11.06 & 1.5 & 6753 &&&&&& \\
      & & & & 11.08 & 1.4 & 11935 &&&&&& \\
\hline
   30 & & & & 11.06 & 1.2 & 2308 &&&&&& \\
      & & & & 11.08 & 1.2 & 15251 &&&&&& \\
\hline 
\end{tabular}\end{center}
\caption{The statistics (in millions of composite sweeps) used in the
 re-weighting analysis for $\beta_c$ in SU($N_c$) gauge theory. Also
 quoted is the integrated auto-correlation time. Studies at $N_t=10$
 and 12 have been carried out with statistics of $4\times10^5$ composite
 sweeps, where $\tau_{int}$ varied between 500 and 1000 sweeps.}
\label{tb.rewt}\end{table}

\begin{table}
\begin{tabular}{|cccc|cccc|}
\hline
\multicolumn{4}{|c|}{SU(4)} & \multicolumn{4}{c|}{SU(6)} \\
\hline
$\beta$ & \multicolumn{3}{c|}{$\langle P\rangle$} & $\beta$ & \multicolumn{3}{c|}{$\langle P\rangle$} \\
\hline
& $N_s=16$ & $N_s=18$ & $N_s=24$ & & $N_s=16$ & $N_s=20$ & $N_s=24$ \\
\hline
10.40 & 0.502033(46) & 0.501991(29) & 0.502136(18) &	24.60 & 0.537686(30) &  &  \\
10.46 & 0.520680(25) & 0.520728(23) & 0.520673(16) &	24.70 & 0.542241(24) &  &  \\
10.48 & 0.525249(12) & 0.525254(16) & 0.525259(7) &	24.80 & 0.546406(9) &  &  \\
10.50 & 0.529260(25) & 0.529268(17) & 0.529274(15) &	24.82 & 0.547189(10) &  &  \\
10.52 & 0.532786(15) & 0.532749(20) & 0.532773(11) &	24.84 & 0.547981(6) &  &  \\
10.54 & 0.535885(9)  & 0.535903(16) & 0.535903(12) &	24.86 & 0.548764(9) &  &  \\
10.60 & 0.543796(10) & 0.543778(12) & 0.543790(7) &	24.88 & 0.549532(8) &  &  \\
10.70 & 0.554088(7) & 0.554097(10) & 0.554092(4) &	24.90 & 0.550292(7) &  &  \\
10.76 & 0.559352(8) &              & 0.559341(5) &	24.92 & 0.551042(6) &  &  \\
10.78 & 0.561000(4) & 0.561005(10) & 0.561000(5) &	25.00 & 0.553961(3) &  &  \\
10.79 &             & 0.561805(8) & 0.561799(4) &	25.10 & 0.557459(5) &  &  \\
10.80 & 0.562606(7) & 0.562605(5) & 0.562602(3) &	25.20 & 0.560805(4) &  &  \\
10.82 & 0.564167(7) & 0.564166(8) & 0.564169(4) &	25.30 & 0.564026(4) & 0.564012(4) & 0.564015(3) \\
10.84 & 0.565700(8) & 0.565689(5) & 0.565702(3) &	25.40 & 0.567126(4) & 0.567119(2) & 0.567120(3) \\
10.90 & 0.570122(5) & 0.570123(4) & 0.570120(3) &	25.42 &             & 0.567727(3) & 0.567727(2) \\
11.00 & 0.576972(5) & 0.576979(5) & 0.576984(3) &	25.44 &             & 0.568332(3) & 0.568329(3) \\
11.02 & 0.578291(5) &             & 0.578286(4) &	25.46 &             & 0.568933(3) & 0.568931(3) \\
11.04 & 0.579577(4) &             & 0.579580(3) &	25.48 &             & 0.569522(3) & 0.569527(3) \\
11.06 & 0.580859(5) &             & 0.580851(4) &	25.50 & 0.570130(5) & 0.570123(4) & 0.570119(3) \\
11.08 & 0.582117(5) &             & 0.582109(4) &	25.52 &             & 0.570709(3) & 0.570708(2) \\
11.10 & 0.583361(5) & 0.583369(3) & 0.583363(3) &	25.54 &             & 0.571293(4) & 0.571294(2) \\
11.20 & 0.589361(4) & 0.589357(3) & 0.589360(2) &	25.56 &             & 0.571880(3) & 0.571877(3) \\
11.30 & 0.595051(3) & 0.595043(4) & 0.595051(2) &	25.60 & 0.573043(3) & 0.573031(3) & 0.573031(2) \\
11.50 & 0.605665(5) & 0.605652(4) & 0.605650(2) &	25.70 & 0.575876(4) &             & \\
11.60 & 0.610655(6) & 0.610643(4) & 0.610630(3) &	25.80 & 0.578631(3) & 0.578620(3) & 0.578616(3) \\
      &             &             &             &	25.90 & 0.581321(4) &             & \\
      &             &             &             &	26.00 & 0.583937(5) & 0.583925(4) & 0.583921(4) \\
      &             &             &             &	26.10 & 0.586506(6) & 0.586487(3) & 0.586485(3) \\
\hline
\end{tabular}
\caption{Plaquette expectation values for SU(4) and SU(6). The numbers in
  brackets denote errors on the least significant digits. Note that the
  volume dependence is negliglible.}
\label{tb.pl46}
\end{table}

\begin{table}
\begin{tabular}{|cc|cc|}
\hline
\multicolumn{2}{|c|}{SU(8)} & \multicolumn{2}{c|}{SU(10)} \\ 
\hline
$\beta$ & $\langle P\rangle$ & $\beta$ & $\langle P\rangle$ \\
\hline
44.50 & 0.542133(9) & 68.00 & 0.376818(6) \\
44.80 & 0.548772(6) & 70.00 & 0.542542(8) \\
45.00 & 0.552860(7) & 71.00 & 0.554874(8) \\
45.20 & 0.556747(5) & 72.00 & 0.567207(8) \\
45.50 & 0.562252(4) & 74.00 & 0.586693(5) \\
46.00 & 0.570739(4) & 76.00 & 0.603381(7) \\
\hline
\end{tabular}
\caption{Plaquette expectation values for SU(8) and SU(10), measured on
  $16^4$ lattices. The numbers in brackets are errors on the least
  significant digit.}
\label{tb.pl810}
\end{table}

\begin{table}
\begin{center}
\begin{tabular*}{0.75\textwidth}{@{\extracolsep{\fill}}cccccc}
\hline
\hline
\multicolumn{3}{c}{SU(4)} & \multicolumn{3}{c}{SU(6)} \\
$\beta$ & $N_t=6$ & $N_t=8$ &  $\beta$ & $N_t=6$ & $N_t=8$ \\
\hline
10.70 & 0.9106(9) & 0.6963(6) &	24.60 & 0.8841(4) & 0.6714(8) \\
10.72 & 0.9311(9) & 0.7119(6) &	24.70 & 0.9335(5) & 0.7089(9) \\
10.74 & 0.9513(10) & 0.7274(6) &	24.80 & 0.9818(5) & 0.7456(9) \\
10.76 & 0.9716(10) & 0.7429(6) &	24.82 & 0.9913(5) & 0.7529(9) \\
10.77 & 0.9817(10) & 0.7507(7) &	24.84 & 1.0010(5) & 0.7602(9) \\
10.78 & 0.9920(10) & 0.7585(7) &	24.85 & 1.0059(5) & 0.7640(9) \\
10.79 & 1.0020(10) & 0.7662(7) &	24.86 & 1.0107(5) & 0.7675(9) \\
10.80 & 1.0122(10) & 0.7740(7) &	24.88 & 1.0204(5) & 0.7749(9) \\
10.82 & 1.0326(10) & 0.7895(7) &	24.90 & 1.0301(5) & 0.7823(9) \\
10.84 & 1.0530(11) & 0.8052(7) &	24.92 & 1.0397(5) & 0.7896(10) \\
10.90 & 1.1151(11) & 0.8526(7) &	25.00 & 1.0786(5) & 0.8191(10) \\
11.00 & 1.2215(12) & 0.9340(8) &	25.10 & 1.1277(6) & 0.8564(10) \\
11.02 & 1.2433(13) & 0.9506(8) &	25.20 & 1.1775(6) & 0.8943(11) \\
11.04 & 1.2654(13) & 0.9675(8) &	25.30 & 1.2282(6) & 0.9327(11) \\
11.06 & 1.2876(13) & 0.9845(9) &	25.34 & 1.2487(6) & 0.9483(11) \\
11.08 & 1.3101(13) & 1.0017(9) &	25.38 & 1.2695(6) & 0.9641(12) \\
11.10 & 1.3331(13) & 1.0193(9) &	25.40 & 1.2799(6) & 0.9720(12) \\
11.12 & 1.3559(14) & 1.0368(9) &	25.42 & 1.2904(6) & 0.9800(12) \\
11.14 & 1.3791(14) & 1.0545(9) &	25.44 & 1.3008(6) & 0.9879(12) \\
11.16 & 1.4028(14) & 1.0726(9) &	25.46 & 1.3114(7) & 0.9960(12) \\
11.18 & 1.4265(14) & 1.0908(10) &	25.48 & 1.3221(7) & 1.0040(12) \\
11.20 & 1.4507(15) & 1.1092(10) &	25.50 & 1.3327(7) & 1.0121(12) \\
11.22 & 1.4748(15) & 1.1276(10) &	25.52 & 1.3434(7) & 1.0202(12) \\
11.24 & 1.4996(15) & 1.1466(10) &	25.54 & 1.3542(7) & 1.0284(12) \\
11.26 & 1.5243(15) & 1.1655(10) &	25.56 & 1.3650(7) & 1.0366(13) \\
11.28 & 1.5496(16) & 1.1849(10) &	25.60 & 1.3868(7) & 1.0532(13) \\
11.30 & 1.5755(16) & 1.2046(10) &	25.80 & 1.4989(7) & 1.1383(14) \\
11.32 & 1.6010(16) & 1.2242(11) &	25.90 & 1.5573(8) & 1.1827(14) \\
11.34 & 1.6271(16) & 1.2441(11) &	26.00 & 1.6169(8) & 1.2280(15) \\
11.36 & 1.6537(17) & 1.2645(11) &	26.20 & 1.7415(9) & 1.3226(16) \\
11.38 & 1.6804(17) & 1.2849(11) &	26.40 & 1.8727(9) & 1.4222(17) \\
11.40 & 1.7077(17) & 1.3058(11) &	26.60 & 2.0119(10) & 1.5279(18) \\
11.42 & 1.7351(17) & 1.3267(12) &	27.00 & 2.3149(11) & 1.7581(21) \\
11.44 & 1.7629(18) & 1.3479(12) &	27.50 & 2.7474(14) & 2.0865(25) \\
11.46 & 1.7911(18) & 1.3696(12) & & & \\	
11.48 & 1.8197(18) & 1.3914(12) & & & \\	
11.50 & 1.8486(19) & 1.4135(12) & & & \\
11.60 & 1.9988(20) & 1.5283(13) & & & \\	
11.70 & 2.1590(22) & 1.6508(14) & & & \\	
11.80 & 2.3300(23) & 1.7816(15) & & & \\	
11.90 & 2.5126(25) & 1.9212(17) & & & \\	
12.00 & 2.7076(27) & 2.0703(18) & & & \\	
12.20 & 3.1390(32) & 2.4001(21) & & & \\	
12.40 & 3.6320(37) & 2.7771(24) & & & \\	
\hline
\end{tabular*}
\end{center}
\caption{$T/T_c$ scales for SU(4) and SU(6) gauge theories in the V-scheme,
 for $N_t=6$ and 8. The errors are dominated by the uncertainty in the
 determination of $\beta_c$.}
\label{tb.tbtc}
\end{table}

\end{document}